# Magnetic domain structure of epitaxial Ni-Mn-Ga films

Anett Diestel[a,b,*], Anja Backen[a,b], Volker Neu[a], Ludwig Schultz[a,b] and Sebastian Fähler[a]

[a] IFW Dresden, Institute of Metallic Materials, P.O. Box 270116, 01171 Dresden, Germany
[b] Dresden University of Technology, Institute of Materials Science, 01062 Dresden, Germany
[*] Corresponding author: a.diestel@ifw-dresden.de, Tel.: +49 (0)351 4659-259

*For the magnetic shape memory effect, knowledge about the interaction between martensitic and magnetic domain structure is essential. In the case of Ni-Mn-Ga bulk material and foils, a staircase-like magnetic domain structure with 90°- and 180°-domain walls is known for modulated martensite. In the present paper we show that the magnetic domain pattern of thin epitaxial films is fundamentally different. Here we analyze epitaxial Ni-Mn-Ga films by atomic and magnetic force microscopy to investigate the correlation between the twinned martensitic variants and the magnetic stripe domains. The observed band-like domains with partially perpendicular out-of-plane magnetization run perpendicular to the microstructure domains defined by twinning variants. These features can be explained by the finite film thickness, resulting in an equilibrium twinning period much smaller than the domain period. This does not allow the formation of a staircase domain patter. Instead the energies of the magnetic and martensitic microstructures are minimized independently by aligning both patterns perpendicularly to each other. By analyzing a thickness series we can show that the observed magnetic domain pattern can be quantitatively described by an adapted band domain model of Kittel.*

# 1 Introduction

Magnetic shape memory (MSM) single crystals reach huge strains up to 10 %, induced by an external magnetic field [1]. This effect is observed in modulated martensite and originates from magnetically induced reorientation (MIR) of differently oriented martensitic variants. The huge strain makes these alloys of particular interest for actuators or sensors in micro-mechanical systems [2]. While the preparation of epitaxial films [3], cantilevers [3,4,5] and freestanding films [6] is well established, little is known about the correlation of magnetic and martensitic microstructure in these microscaled systems. Since coupling of magnetic and martensitic microstructure is the key requirement for MIR, a better understanding of possible finite size effects is necessary for the functionalization of these microstructures. For this, bulk single crystals [7,8,9] and foils [10] can be considered as a reference system. The variants in multivariant state of bulk Ni-Mn-Ga typically run through the complete sample. Within one single variant the formation of antiparallel domains separated by 180°-domain walls (DW) is observed. 90°-DW form at twin variant boundaries and enable the magnetization to follow the magnetic easy *c*-axis. With the formation of a typical staircase pattern consisting of 90°- and 180°-DW the magnetic flux can be closed [7-10] (see Fig. 1c). In addition to this fundamental pattern domain mirroring at twin boundaries [11] and branching at slightly miscut surfaces [12] have been analyzed in single crystals.

Until now the magnetic domain pattern of thin films has only been analyzed in polycrystalline Ni-Mn-Ga [13]. Magnetic force microscopy micrographs reveal a strong magnetic out-of-plane contrast produced by band domains forming a maze like pattern. A thickness series revealed that the domain period $\Lambda_{DW}$ depends on the film thickness $d$ as predicted by Kittel ($\Lambda_{DW} \sim d^{1/2}$) [14]. However, the physical origin of the prefactor in Kittel's relation is missing and due to the polycrystalline nature of these films a correlation of magnetic domains and martensitic variants was not possible [13].

In the present paper we use the advantage of epitaxial films, which allow a local identification of the martensitic microstructure by atomic force microscopy (AFM) [15,16]. Magnetic force microscopy (MFM) measurements reveal a correlation of magnetic and martensitic microstructure of 14M martensite, which is fundamentally different compared to bulk. We show that these differences originate from the finite film thickness and the reduced martensitic variant periodicity. We analyze the thickness dependency of the magnetic domain period and compare it with the qualitative prediction of Kittel's theory for magnetic band domains [14]. Furthermore we demonstrate that a modified band domain theory leads to a full quantitative explanation of the thickness dependency without any free parameter.

Our domain model benefits from the fact, that Ni-Mn-Ga is a well characterized system. Direct measurements of the spin wave stiffness constant $W = 100$ meVÅ$^2$ [17] allow calculating the exchange constant $A = \frac{W \cdot S \cdot N}{2a^3}$ by using the atomic moment $S = 3.8\mu_B$, the number $N$ of Mn atoms per unit cell and the corresponding lattice parameter $a$ of Ni-Mn-Ga. The uniaxial anisotropy coefficient for the present 14M martensite is $K_u = 0.9 \cdot 10^5$ Jm$^{-3}$ [18].

## 2 Experimental

We prepared a series of epitaxial Ni-Mn-Ga films with thicknesses $d$ ranging from 125 nm to 2 µm by DC magnetron sputtering, as described in detail in our previous work [15]. In order to improve the film quality and enable freestanding Ni-Mn-Ga films, we first deposited a sacrificial chromium layer on the single crystalline MgO(100) substrate [6]. Film composition was probed by means of energy-dispersive X-ray spectroscopy (EDX) with an accuracy of 0.5 at.% using a Ni$_{50}$Mn$_{25}$Ga$_{25}$ standard. The valence electron density (*e/a*-ratio) was determined to be about 7.61 (±0.06) by these measurements. Martensitic and magnetic microstructure were investigated by AFM and MFM, respectively, using a digital instrument dimension 3100. Topography was imaged by height contrast in tapping mode. Magnetic micrographs were obtained in lift mode with a standard magnetic tip (Co-alloy coating, magnetization along tip axis). The lift scan height ranges from 50 to 100 nm depending on the film thickness.

## 3 Results and Discussion

### 3.1 Magnetic domain configuration in 14M martensite

Crystallographic and magnetic microstructures were measured by AFM and MFM and are exemplarily shown for a 2 µm thick Ni-Mn-Ga film in Fig. 1. Due to the epitaxial film growth, the crystallographic orientation of the austenitic Ni-Mn-Ga unit cell is known [15]. It is rotated about 45° with respect to the substrate and image edges. The AFM micrograph in Fig. 1a shows a periodically modulated waved topography with a rhombus-like superstructure. This pattern represents traces of mesoscopic *c-a*-twin boundaries of the modulated 14M martensitic phase [15]. Due to the four-folded symmetry of the epitaxial (001) oriented film two crystallographic equivalent orientations of twin boundary traces along MgO[011] and MgO[0$\bar{1}$1] are possible. In both cases the traces of twin boundaries are rotated about 45° with respect to the substrate edges [16].

With the knowledge about twinned 14M martensite and the presented AFM micrographs the crystallography can be sketched schematically as top view and cross section in Fig. 1b. The diagonal, black lines illustrate the traces of mesoscopic $c_{14M}$-$a_{14M}$-twin boundaries (TB) and the double arrows are equivalent to the crystallographic short and magnetic easy $c_{14M}$-axis. The top view displays the alternating in- and out-of-plane $c_{14M}$-axes of adjoining martensitic variants. The cross section cut along MgO[011] direction (Fig. 1b, dotted line) sketches the tilted run of the twin boundaries within the film.

By transferring the magnetic domain structure of bulk Ni-Mn-Ga [7] onto the given arrangement of martensitic variants a proposed domain structure can be sketched in Fig. 1c. The $c_{14M}$-axis is equivalent to the preferential direction of magnetization $\vec{m}$, which is illustrated by red arrows. According to the findings on bulk Ni-Mn-Ga, the cross section along MgO[011] should exhibit a staircase domain pattern of 90°- and 180°-domain walls (DW, red dashed lines). 90°-DW coincide with the tilted twin boundaries, whereas 180°-DW separate oppositely magnetized domains within one martensitic variant. This typical structure minimizes the magnetic stray field and closes the magnetic flux within the sample. The staircase domain structure inside the sample would cause a domain pattern on the film surface as sketched in the top view (Fig. 1c). A band like out-of-plane contrast along the twin boundaries should be observed by MFM.

In contrast to these expectations, the experimentally observed domain pattern (Fig. 1d) presents a lamellar bright-dark-contrast perpendicular to the twin boundaries, which suggests the formation of magnetic band domains with antiparallel out-of-plane magnetization. Within the dark domains a finer contrast is visible. This contrast coincides exactly with the topography contrast in the AFM micrograph, which cannot be excluded completely during MFM measurements for larger film thicknesses ($d \geq 1$ µm). Due to branching and bending the magnetic domains do not run perfectly parallel to each other, though we could not find any correlation between topographic features and these discontinuities.

In Fig. 1e the entire data obtained from AFM and MFM micrographs are summed up. The schematic top view and cross section explain the visible contrast by means of a possible domain configuration. From AFM measurements the crystallography and the orientation of twin boundaries and $c_{14M}$-axes are known. For the interpretation of the magnetic configuration we consider that the magnetization $\vec{m}$ follows the magnetic easy $c_{14M}$-axis in every segment of the band domains, which alternates from in- to out-of-plane between neighboring variants. In the following we discuss the possible alignments of magnetization (up, left, down, right) within the magnetic domains. Instead of permutation of all four possible magnetization directions within each band domain, only two magnetic directions occur within each band domain (e. g. up and left in the dark band domain, Fig. 1e cross section). In the case of not-observed permutation of all four directions every second 90°-DW would be charged (Fig. 1f). So this configuration is energetically unfavorable.

We observe only little magnetic contrast within each band domain. This indicates for coupling effects, which brings magnetization even more parallel. In particular magnetostatic coupling favors a more parallel alignment of magnetization within one band domain, which is illustrated by the slightly tilted magnetizations in the cross section of Fig. 1e. The same parallel alignment would be also favored by exchange coupling across twin boundaries. However, the exchange length with $L_{ex} = \sqrt{A \cdot K_u^{-1}} = 8.3$ nm is quite low in comparison to the twin variant width in our films. Exchange coupling therefore should only play a role in very thin films with a short twinning period.

Coupling results in an out-of-plane component of magnetization. To minimize stray field energy antiparallel magnetized band domains form, which are divided by 180°-DW (Fig. 1e). For band domains the equilibrium domain period $\Lambda_{DW}$ is an optimum balance between total domain wall energy, which increases with $\Lambda_{DW}$, and stray field energy, which reduces with $\Lambda_{DW}$ [14]. As we will describe in detail in 3.3, the experimentally observed $\Lambda_{DW}$ agrees well with the calculated one. Since $\Lambda_{DW}$ is substantially larger than the twinning period $\Lambda_{TB}$ the formation of domains along the direction of the twin boundaries does not allow reaching their optimum period. For a domain wall orientation perpendicular to the twin boundaries this restriction does not exist and every domain width can be formed to adjust the equilibrium domain configuration. This magnetic domain structure is visualized by MFM as bright and dark stripes perpendicular to the twin boundaries (see Fig. 1d).

To understand why this domain configuration is more favorable in thin films compared to the bulk staircase pattern we estimate the domain wall energies in both configurations (Fig 1c, e). When comparing the total domain wall energy, it is sufficient to compare the areas of 180°-DWs only, since in both cases the area of the 90°-DW is identical and determined by the twin boundary area. For the staircase pattern sketched in Fig. 1c the ratio of DW area per film surface area is at least $2d/\Lambda_{TB} = 10.6$ (or a multiple) for the presented 2 µm thick film. For the experimentally observed magnetic band domain pattern this ratio depends on film thickness $d$ (since the DWs are perpendicular to the substrate) and their period $\Lambda_{DW}$. The ratio $2d/\Lambda_{DW}$ gives a value of 3.3. This illustrates that the total domain wall energy is substantially lower for band domains compared to a staircase domain pattern.

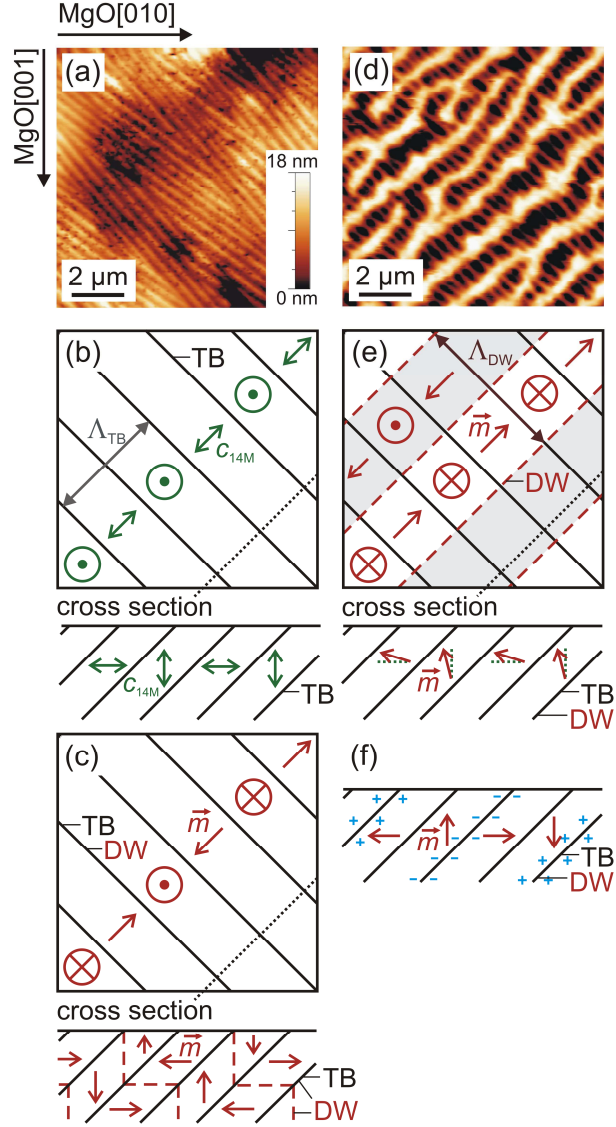

FIG. 1. (Color online) (a) The AFM micrograph of a 2 µm thick Ni-Mn-Ga film shows diagonal traces of mesoscopic $c_{14M}$-$a_{14M}$-twin boundaries. (b) The crystallographic orientation is sketched as top view and cross section, where the black lines illustrate the twin boundaries (TB) with the twinning period $\Lambda_{TB}$ and the double arrows picture the crystallographic short and magnetic easy $c_{14M}$-axis. (c) Following the known magnetic domain pattern of bulk materials and the crystallography from the AFM image, a hypothetical magnetic domain structure can be suggested. For the cross section a staircase pattern of 180°-domain walls (DW, red dashed lines) is sketched, where the magnetization (arrows) follows the $c_{14M}$-axis and the 90°-DW coincide with the TB. (d) In contrast to this, the experimental MFM micrograph shows a high out-of-plane lamellar contrast perpendicular to the TB. (e) The corresponding domain structure with the domain width period $\Lambda_{DW}$ is sketched schematically as top view and cross section. (f) The cross section shows the energetically unfavourable, experimentally not observed domain configuration with charged 90°-DWs.

## 3.2 Film thickness dependency

The change of the magnetic domain pattern as a function of the film thickness $d$ in the range of 125 nm to 2 µm was determined by analyzing MFM micrographs shown in Fig. 2. The corresponding AFM images are not shown here, but they reveal a similar topography caused by mesoscopic $c_{14M}$-$a_{14M}$-twin boundaries imaged in Fig. 1a.

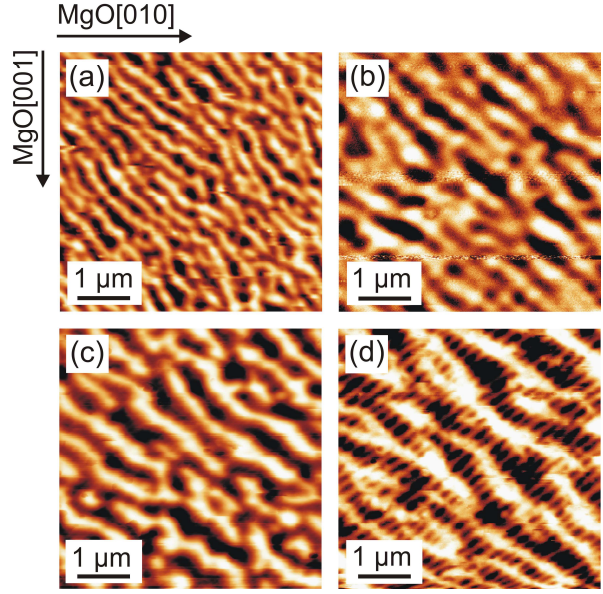

FIG. 2. (Color online) MFM micrographs of Ni-Mn-Ga films with thicknesses $d$ of (a) 125 nm, (b) 250 nm, (c) 500 nm and (d) 1 µm show an increasing domain width period $\Lambda_{DW}$. For $d \geq 1$ µm fine lines originating from the topographic contrast of twin boundaries become visible.

In analogy to magnetic band domains proposed by Kittel [14] the equilibrium twinning period can be described by minimization of the sum of elastic volume energy and twin boundary energy. Accordingly the twinning period can be described with the power law $\Lambda_{TB}(d) = a \cdot d^{\nu}$. A non-linear fitting of $\Lambda_{TB}$ versus $d$ results in the parameters $\nu = 0.87(\pm 0.04)$ and $a = 0.53(\pm 0.16)$ nm$^{1-\nu}$ (Fig. 3). In our previous work we analyzed a similar film series and obtained a different parameter $\nu = 1/2$ [16]. The exponent allows distinguishing different solutions to minimize both, twin boundary energy and elastic energy [19]. While $\nu = 1/2$ describes a constant twin boundary density through the film thickness, $\nu = 2/3$ is expected when branching of twin boundaries occurs. The clear difference between both film series becomes also evident by different shapes of the mesoscopic $c_{14M}$-$a_{14M}$-twins on the surface. The present topography shows a sharp rhombus-like superstructure whereas the previous one exhibits a meandering pattern [16]. This indicates a different relaxation mechanism, which will be analyzed in detail elsewhere. In the present paper we take the twin boundary periodicity as given and focus on the correlation with the magnetic domain pattern.

For the magnetic domain structure the non-linear fitting of the domain width period $\Lambda_{DW}(d) = a \cdot d^\nu$ results in the parameters $\nu = 0.46(\pm 0.04)$ and $a = 37.1(\pm 8.7)$ nm$^{1-\nu}$ (Fig. 3). This demonstrates that the observed magnetic domain structure obeys the theoretical thickness dependency of magnetic band domains predicted by Kittel [14]. By fixing $\nu = 1/2$ the dependency of $\Lambda_{DW}(\text{nm}) = 27.7$ nm$^{1/2} \cdot d^{1/2}(\text{nm}^{1/2})$ is obtained. To estimate the thickness for which the magnetic domain pattern should change from the thin film configuration towards the bulk staircase pattern the different thickness dependencies of domain and variant periodicity is used. A crude extrapolation of both cases depicted in Fig. 3 suggests a crossover for film thicknesses in the millimeter range.

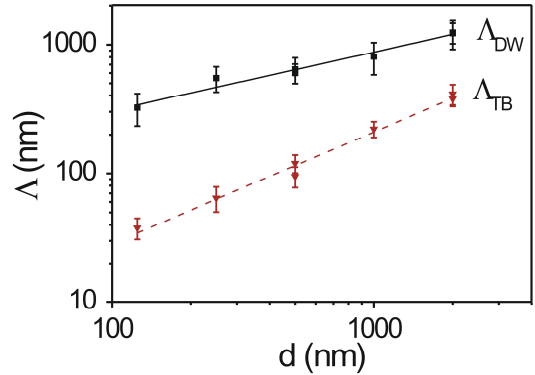

FIG. 3. (Color online) Double-logarithmic plots of periodicity versus film thickness $d$ to determine the exponent $\nu$ in the film thickness dependency $\Lambda(d) = a \cdot d^\nu$ is shown. For the twinning period $\Lambda_{TB}$ (triangle) an exponent of $\nu = 0.87 \pm 0.04$ is obtained. For the domain width period $\Lambda_{DW}$ (square) an exponent of $\nu = 0.46 \pm 0.04$ is obtained, which allows using a square-root thickness dependency $\Lambda_{DW}(\text{nm}) = 27.7$ nm$^{1/2} \cdot d^{1/2}(\text{nm}^{1/2})$.

### 3.3 Quantitative description according to the Kittel model

Materials with an uniaxial anisotropy and out-of-plane magnetic easy axis often show a band domain structure. This domain configuration was theoretically studied by Kittel and applies for materials with a quality-factor $Q = \frac{K_u}{K_d}$ larger than one [14]. $K_d = \frac{J_S^2}{2\mu_0}$ denotes the magnetostatic energy density with the saturation magnetization $J_S = 0.6$ T. Under these conditions $Q$ is 0.64, which is smaller than one and the application of the Kittel model is not valid.

As described in 3.1 magnetostatic coupling results in an averaged, tilted magnetization. Then the effective saturation magnetization ($J_S^*$) is reduced and as projection to the film normal $J_S^* = \frac{1}{\sqrt{2}} J_S$ is obtained (Fig. 4a). In this case the corresponding quality factor $Q^*$ is 1.29, which allows the comparison with the band domain theory.

According to Kittel the band domain structure is formed by minimizing the total energy as sum of magnetic domain wall and stray field energy. The equilibrium domain width $D$ can be determined by

$$D = \sqrt{\frac{\pi^3 \cdot \gamma_w \cdot d}{8 K_d^*}} . \qquad (1)$$

With 180° Bloch domain walls exhibiting an energy of $\gamma_w = 4\sqrt{A \cdot K_u}$, the film thickness dependence of the domain width can be determined with Eq. 1. The domain width periodicity $\Lambda_{DW}$ represents the sum of two adjoining domain widths $D$ and was calculated: $\Lambda_{DW}(\text{nm}) = 25.8 \text{ nm}^{1/2} \cdot d^{1/2}(\text{nm}^{1/2})$. Fig. 4b shows the relation of the experimentally determined domain width period (squares) and the calculated $\Lambda_{DW}$ (graph) as function of the film thickness $d$. The deviation of the experimental data from the calculated function is only minor and may be due to the simplification of the complex magnetic domain structure. The high conformity between experimental data and theory indicates that no additional domain walls are embedded parallel to the film surface.

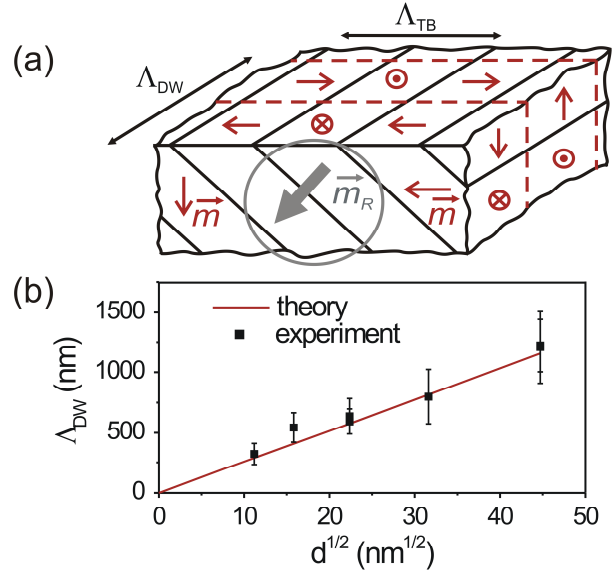

FIG. 4. (Color online) (a) Three dimensional sketch images the simplified magnetic domain structure of thin epitaxial Ni-Mn-Ga films. The short twinning period $\Lambda_{TB}$ results in exchange coupling of the magnetization $\vec{m}$ between adjoining variants. The resulting exchange coupled magnetization $\vec{m}_R$ exhibits an out-of-plane component. To reduce their stray field energy band domains with the periodicity $\Lambda_{DW}$ form. (b) Comparison of experimentally determined domain width period $\Lambda_{DW}$ (black squares) and calculated values from Kittel's theory of magnetic band domains (red line) as a function of film thickness $d$.

## 4 Conclusions

The finite film thickness and twin variant width does not allow the formation of a staircase pattern known for Ni-Mn-Ga bulk materials. Due to the high aspect ratio of thin films, stray field energy has an impact on the magnetic domain pattern and band domains are formed. Their microscopically analyzed width can be described quantitatively without any free parameter by Kittel's theoretical model of magnetic band domains. In the analysed thin films the equilibrium magnetic band domain period exceeds the twin boundary period. In agreement with our experiments the energies of magnetic as well as martensitic microstructure can be minimized independently, when domain and twin boundaries are aligned perpendicular to each other.

## 5 Acknowledgments

We acknoledge Rudolk Schäfer and Ulrich K. Rößler for helpful discussions. This work was funded by the German research Foundation (DFG) via the Priority Programme SPP1239.